\documentclass[prd]{revtex4}

\usepackage{graphicx}
\usepackage{amsmath}
\usepackage{amssymb}
\usepackage{dcolumn}

\newcommand{\Slashed}[1]{\ensuremath{{#1}{\!}{\!}{\!}{\!}{\!}{\:}/}}

\newcommand{\prettyfraction}[2]{\ensuremath{{{}^{#1}{\!}/{\!}{}_{#2}}}}
\newcommand{\eg}[0]{\textit{e.g.}}
\newcommand{\ie}[0]{\textit{i.e.}}

\newcommand{\citereference}[1]{Ref.~\cite{#1}}
\newcommand{\citefigure}[1]{Fig.~\ref{#1}}
\newcommand{\citetable}[1]{Tab.~\ref{#1}}
\newcommand{\citesection}[1]{Sec.~\ref{#1}}

\begin{document}

\title{Single B Production through R-Parity Violation}

\author{Ben O'Leary}
\affiliation{SUPA, School of Physics, University of Edinburgh, Edinburgh EH9 3JZ, Scotland, U.K.}
\email{ben.oleary@ed.ac.uk}

\date{\today}

\begin{abstract}
Supersymmetry without $R$--parity predicts tree level quark flavor violation.  We present a potential signal of single bottom production at electron--positron colliders with energies in the range $6$ to $20$~GeV.  Taking into account rare decay limits, it should be detectable with the current BaBar and Belle data samples.
\end{abstract}


\maketitle


\section{Introduction}

The Minimal Supersymmetric Standard Model~\cite{MSSM_review} (MSSM) without $R$--parity~\cite{RPV_review} predicts Yukawa interactions that violate baryon and/or lepton number without violating Standard Model gauge symmetries or supersymmetry.  The superpotential for these interactions is
\begin{eqnarray}
W_{{\Slashed{R}}} & = &   {\prettyfraction{1}{2}} {\lambda}_{ijk} ( L_{i} )_{a} {\epsilon}_{ab} ( L_{j} )_{b} E_{k}^{c}
                        + {{\lambda}'}_{ijk} ( L_{i} )_{a} {\epsilon}_{ab} ( Q_{j} )_{b} D_{k}^{c} \nonumber\\
                  &   & + {{\mu}'}_{i} ( L_{i} )_{a} {\epsilon}_{ab} ( H_{u} )_{b}
                        + {\prettyfraction{1}{2}} {{\lambda}''}_{ijk} U_{i}^{c} D_{j}^{c} D_{k}^{c}
\end{eqnarray}
where $i$, $j$ and $k$ are generational indices and $a$ and $b$ are $SU(2)$ indices.  Color indices have been suppressed. $L$ is the lepton doublet, $E^{c}$ is the charged anti--lepton singlet, $Q$ is the quark doublet, $D^{c}$ is the down--type anti--quark singlet, $U^{c}$ is the up--type anti--quark singlet and $H_{u}$ is the Higgs doublet which generates mass for up-type quarks.  The factor of ${\prettyfraction{1}{2}}$ before the ${\lambda}$ term is conventional: we see that ${\lambda}_{ijk} = -{\lambda}_{jik}$ by relabelling $a$ to $b$ and vice--versa in the first term of the superpotential, hence the extra factor of ${\prettyfraction{1}{2}}$ sets the coupling of the ${\nu}_{i} e_{j} E_{k}^{c}$ term to be ${\lambda}_{ijk}$ rather than $2 {\lambda}_{ijk}$.  Likewise ${{\lambda}''}_{ijk}$ is antisymmetric in $j$ and $k$ (the color indices in the final term are combined with an antisymmetric tensor), hence its factor of ${\prettyfraction{1}{2}}$.

The combination of ${{\lambda}'}$ and ${{\lambda}''}$ leads to proton decay and is thus constrained by searches for proton decay into a positron and a pion~\cite{proton_decay_to_e_pi_ref} and also by invisible neutron disappearance searches~\cite{neutron_decay_to_invisible_ref}.  Requiring ${{\lambda}''}_{ijk} = 0$ is sufficent to guarantee perturbative proton stability while leaving the possibility of non--zero lepton--number violating couplings.  These couplings introduce a new channel for flavor violation.

So far, there is no direct evidence for supersymmetry or $R$--parity violation.  The non--observation of single sparticle production puts constraints on a combination of their masses and couplings~\cite{PDG}.  Most of the tightest bounds on individual couplings come from charged current universality~\cite{charged_current_universality}, as single sfermion exchange generally interferes with weak boson exchange.  Meson and ${\tau}$--lepton rare decay data typically provide tighter bounds on products of $R$--parity violating (RPV) couplings than the product of individual bounds~\cite{RPV_review}.

In an era of high--precision flavor physics the obvious question is whether for example $B$ physics observables can be used to further probe the RPV parameter space.  We present a potential signal of single $B$ production at electron--positron colliders with energies in the range $6$ to $20$~GeV, with special attention given to the case of a center--of--mass energy of $10.58$~GeV, at which BaBar and Belle currently run.  The lower limit is chosen slightly above the threshold for creating a $B K$ meson pair.

This paper is arranged as follows.  First, the potential signal through RPV couplings is calculated for two cases: that in which only an on--shell quark---anti--quark pair is produced, and that in which an on--shell quark---anti--quark pair and a photon are produced.  The hadronization of the quarks and the experimental signature are then briefly discussed.  Next, the background to the signal is considered: the SM contribution is calculated and the signal from the $R$--parity-conserving sector of the MSSM is estimated.  Finally, the conclusions are presented.

\section{Single ${\mathbf{b}}$ Production}

The Standard Model predicts quark flavor violation through CKM mixing, but the cross--sections for $e {\bar{e}} {\to} b {\bar{s}}$ or $b {\bar{d}}$ are extremely small.  Detection of flavor violation in significant excess to the Standard Model prediction would be an exciting signal of new physics.

In the MSSM without $R$--parity, flavor violation can be mediated at tree level by sfermions.  A non--zero ${\lambda}$ and ${{\lambda}'}$ combination allows single $b$ quark production along with a light down--type anti--quark through the Feynman diagrams shown in \citefigure{single_b_RPV_diagrams}.

\begin{figure}[t] 
\begin{center}
\leavevmode
\includegraphics[width=10cm]{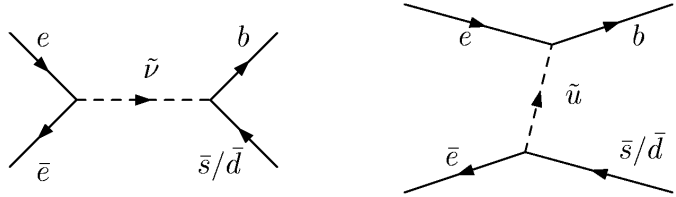}
\caption{\label{single_b_RPV_diagrams}Sneutrino--  and squark--mediated single $b$ production diagrams.}
\end{center}
\end{figure}

The sneutrino mediated diagram is proportional to ${\lambda}_{g11}^{{\ast}}$ ($e {\bar{e}} {\tilde{\nu}}$ vertex) multiplied by either ${{\lambda}'}_{g23}$ (${\tilde{\nu}} b {\bar{s}}$ vertex) or ${{\lambda}'}_{g13}$ (${\tilde{\nu}} b {\bar{d}}$ vertex).  The squark mediated diagram is proportional to ${{\lambda}'}_{1g3}$ ($e b {\tilde{u}}$ vertex) multiplied by either ${{\lambda}'}_{1g2}^{{\ast}}$ (${\tilde{u}} {\bar{e}} {\bar{s}}$ vertex) or ${{\lambda}'}_{1g3}$ (${\tilde{u}} {\bar{e}} {\bar{d}}$ vertex).  We write ${\tilde{\nu}}$ and ${\tilde{u}}$ instead of ${\tilde{\nu}}_{L}^{g}$ and ${\tilde{u}}_{L}^{g}$ to avoid clutter --- in all cases the sfermion is implicitly of generation $g$, and associated with the left--handed chirality of its superfield partner.

Because the sfermions are constrained to be heavy, $m_{{\tilde{\nu}},{\tilde{u}}} {\gtrsim} 100$~GeV~${\gg} {\sqrt{s}}$, we approximate their propagators as static $1/m_{{\tilde{\nu}},{\tilde{u}}}^2$.  Moreover, we assume that one sfermion dominates the signal process, either because it is lighter than the others or because it has a larger coupling product.

If we sum over the $b$ and ${\bar{b}}$ final states and allow for only one of the two $R$--parity violating processes to dominate we obtain the differential cross--sections
\begin{equation}
{\frac{d{\sigma}}{d{\Omega}}} = {\frac{|{\mathbf{p}}_{b}|}{|{\mathbf{p}}_{e}|}} \;
                            {\frac{3}{128 {\pi}^{2}}} \; 
                            \left( s - m_{b}^{2} - m_{s}^{2} \right) \;
                            {\frac{|{\lambda}_{g11}|^{2} |{{\lambda}'}_{g23}|^{2}}{{m_{{\tilde{{\nu}}}}^{4}}}}
\label{snu_d_sigma_d_Omega_equation}
\end{equation}
for the $s$--channel sneutrino exchange, and
\begin{equation}
{\frac{d{\sigma}}{d{\Omega}}} = {\frac{|{\mathbf{p}}_{b}|}{|{\mathbf{p}}_{e}|}} \;
                            {\frac{3}{128 {\pi}^{2} s}} \;
                            \left( t - m_{b}^{2} \right) \;
                            \left( t - m_{s}^{2} \right) \;
                            {\frac{|{{\lambda}'}_{1g2}|^{2} |{{\lambda}'}_{1g3}|^{2}}{{m_{{\tilde{u}}}^{4}}}}
\label{squ_d_sigma_d_Omega_equation}
\end{equation}
for the $t$--channel squark.  ${\mathbf{p}}_{b}$ is the 3--momentum of the $b$ quark. We have ignored the electron mass compared to the rest of the masses and energies.  The case of a final--state down quark can be obtained by the appropriate changes of indices.  The current limits on these combinations of couplings from experimental data~\footnote{We note that \citereference{my_RPV_bounds_paper}, which appeared while this work was under review, presents corrected bounds from snuetrino--mediated $B$--decay which disagree with those from \citereference{Saha_Kundu}, by being less tight.  However, it is only by a factor of roughly $6$, which is not enough to raise the signal beyond the attobarn level.} are given in \citetable{RPV_decay_bounds_table}.

\begin{table}[t]
\begin{center}
\begin{tabular}{|c|dl|l|}
\hline
Coupling & \multicolumn{2}{c}{Bound} & Process \\
\hline
$|{\lambda}_{g11}|^{2} |{{\lambda}'}_{g13}|^{2} m_{{\tilde{{\nu}}}}^{-4}$ & 2.9 & ${\times} 10^{-18}$~GeV${}^{-4}$ & $B_{d}^{0} {\to} e {\bar{e}}$~\cite{Saha_Kundu}\\
\hline
$|{\lambda}_{g11}|^{2} |{{\lambda}'}_{g23}|^{2} m_{{\tilde{{\nu}}}}^{-4}$ & 5.9 & ${\times} 10^{-18}$~GeV${}^{-4}$ & $B_{s}^{0} {\to} e {\bar{e}}$~\cite{Xu_Wang_Yang}\\
\hline
$|{{\lambda}'}_{1g1}|^{2} |{{\lambda}'}_{1g3}|^{2} m_{{\tilde{u}}}^{-4}$  & 2.9 & ${\times} 10^{-13}$~GeV${}^{-4}$ & APV in Cs~\cite{APV_in_Cs}, $A_{\text{FB}}^{b}$~\cite{RPV_review}\\
\hline
$|{{\lambda}'}_{1g2}|^{2} |{{\lambda}'}_{1g3}|^{2} m_{{\tilde{u}}}^{-4}$  & 2.2 & ${\times} 10^{-17}$~GeV${}^{-4}$ & $B_{s}^{0} {\to} K e {\bar{e}}$~\cite{Xu_Wang_Yang}\\
\hline
\end{tabular}
\end{center}
\caption{\label{RPV_decay_bounds_table}Bounds on coupling combinations.  The atomic parity violation bound on $|{{\lambda}'}_{1g1}|^{2} m_{{\tilde{u}}}^{-2}$ is combined with the constraint on $|{{\lambda}'}_{1g3}|^{2} m_{{\tilde{u}}}^{-2}$ from the bottom forward--backward asymmetry.}
\end{table}

In calculating the signal, we assume that the values of the couplings are equal to their current bounds.  Performing the angular integrations (restricted to $|{\cos}( {\theta} )| {\leq} 0.9$) leads to the cross--sections presented in \citefigure{sigma_down_plot} and \citefigure{sigma_strange_plot}, with the numerical values for ${\sqrt{s}} = 10.58$~GeV given in \citetable{RPV_cross-section_values}.

\begin{table}[t]
\begin{center}
\begin{tabular}{|r|dr||r|dr|}
\hline
$bd$ via ${\tilde{{\nu}}}$ & 3.4  & ${\times} 10^{-6}$~fb & $bd{\gamma}$ via ${\tilde{{\nu}}}$ & 1.6 & ${\times} 10^{-9}$~fb\\
\hline
$bs$ via ${\tilde{{\nu}}}$ & 6.2  & ${\times} 10^{-6}$~fb & $bs{\gamma}$ via ${\tilde{{\nu}}}$ & 2.9 & ${\times} 10^{-9}$~fb\\
\hline
$bd$ via ${\tilde{u}}$     & 0.13 & fb                    & $bd{\gamma}$ via ${\tilde{u}}$     & 5.7 & ${\times} 10^{-5}$~fb\\
\hline
$bs$ via ${\tilde{u}}$     & 1.0  & ${\times} 10^{-5}$~fb & $bs{\gamma}$ via ${\tilde{u}}$     & 4.3 & ${\times} 10^{-9}$~fb\\
\hline
\end{tabular}
\end{center}
\caption{\label{RPV_cross-section_values}The cross--sections for $e {\bar{e}} {\to} b {\bar{s}} / s {\bar{b}} / b {\bar{d}} / d {\bar{b}}/ b {\bar{s}} {\gamma} / s {\bar{b}} {\gamma} / b {\bar{d}} {\gamma} / d {\bar{b}} {\gamma}$ at ${\sqrt{s}} = 10.58$~GeV.}
\end{table}

\subsection{Single ${\mathbf{b}}$ Production With A High--Energy Photon}

As is discussed in \citesection{Background_section}, the production of a single $B$ meson --- light meson pair is not necessarily a clean signal.  $B$ mesons are often misidentified, and an accurate reconstruction of the kinematics may reduce the detection efficiency substancially.  Here we consider the cases of an additional final--state photon for the signals considered above, which may prove to be a cleaner signal as the energy of the $B$ meson does not have to be measured --- for a sufficiently energetic photon, $B {\bar{B}}$ pair production is kinematically excluded (in analogy to using radiative return to measure hadronic cross--sections for lower energies than those at which an experiment runs~\cite{radiative_return}).  The Feynman diagrams are the same as in \citefigure{single_b_RPV_diagrams}, but with an external photon emitted by any of the external particles.  An emission by the virtual squark suppresses the matrix element by another power of $m_{{\tilde{u}}}^{2}$.  We make a restriction on the photon to exclude the possibility that it was emitted through the radiative decay of a $B$ meson.  Since there is an upper bound to the energy that the radiated photon can have for a $B$ meson with a given momentum in the beam center--of--momentum frame, we restrict the photon to have $10\%$ or more energy above this value for a $B$ meson with half the beam energy, \ie
\begin{equation}
E_{{\gamma}} {\geq} 1.1 {\frac{{m_{B}^{2}}}{2 ( ( {\sqrt{s}} / 2 ) - {\sqrt{s / 4 - m_{B}^{2}}} )}}
\end{equation}
In doing this, we eliminate the background of misidentified $B {\bar{B}}$ pair production.

The cross--sections for this process are also presented in \citefigure{sigma_down_plot} and \citefigure{sigma_strange_plot}, with the numerical values for ${\sqrt{s}} = 10.58$~GeV given in \citetable{RPV_cross-section_values}.  The signal begins at $10.56$~GeV as below this it is kinematically impossible to produce a $B {\bar{B}}$ pair, hence the advantage of the additional photon is non--existent, while still suffering from the ${\alpha}$ suppression of the signal.  The restriction on the photon energy cuts out much of the phase space, and cuts out more as ${\sqrt{s}}$ increases, until around ${\sqrt{s}} = 13.8$~GeV, where the entire phase space is excluded.  Unfortunately, even in the best case, close to the special value ${\sqrt{s}} = 10.58$~GeV, the best signal is less than $0.1$~ab.

\subsection{Experimental Signature}

The signal calculated above has on--shell single quarks in the final state.  The process of hadronization is not well understood, but since ${\sqrt{s}} {\gg} {\Lambda}_{{\text{QCD}}}$ we assume that the scattering amplitude for the sum of all possible $e {\bar{e}} {\to} M {\bar{B}}$ is the same as for $e {\bar{e}} {\to} b {\bar{q}}$, where $M$ is a light (bottomless) meson which has anti--quark constituent ${\bar{q}}$.  In this scheme, the production of a $b {\bar{d}}$ pair leads to an on--shell neutral pair with an unflavored light meson (${\bar{B}}_{d} {\pi}^{0}$, ${\bar{B}}_{d} {\eta}$, ${\bar{B}}_{d} {{\eta}'}$, ${\bar{B}}_{d}^{{\ast}} {\rho}$ or ${\bar{B}}_{d}^{{\ast}} {\omega}$) $43.5\%$ of the time and to a charged pair with an unflavored light meson ($B_{d}^{-} {\pi}^{+}$, $B_{d}^{{\ast}-} {\rho}^{+}$) $43.5\%$ of the time, according to the Lund string model~\cite{Lund_string}. The remaining $13\%$ consist of the channels where the light meson is strange (${\bar{B}}_{s} K$ and ${\bar{B}}_{s}^{{\ast}} K^{{\ast}}$).

\section{Background}
\label{Background_section}

We identify three sources of background to the signal: direct SM $e {\bar{e}} {\to} M {\bar{B}}$, misidentified $B {\bar{B}}$ pair production, and $R$--parity conserving MSSM $e {\bar{e}} {\to} b {\bar{s}}$ or $b {\bar{d}}$.

\subsection{Standard Model Background}

As mentioned in the introduction, there is a Standard Model background to the processes $e {\bar{e}} {\to} b {\bar{s}}, b {\bar{d}}$.  However, its leading order contribution is at one--loop level and is Cabbibo suppressed.  Ignoring Feynman diagrams with a electron--Higgs Yukawa coupling, there are five classes of diagrams, shown in \citefigure{single_b_SM_2q_diagrams} (in these diagrams the photon may be replaced by a $Z$ boson, though this suppresses the matrix element by a further factor of $s / m_{Z}^{2}$).

\begin{figure}[t]
\begin{center}
\leavevmode
\includegraphics[width=12cm]{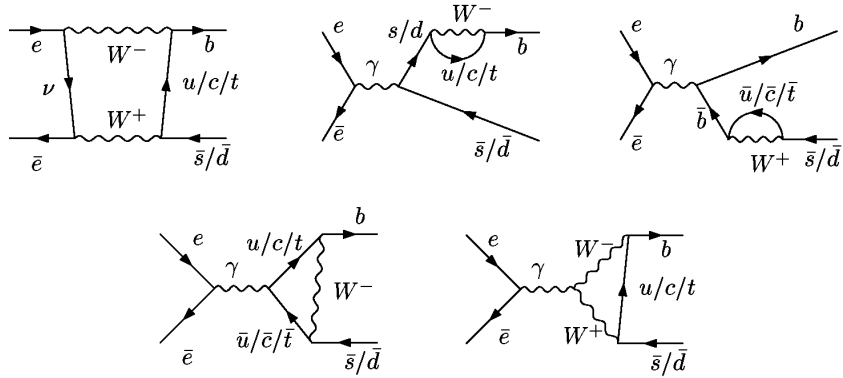}
\caption{\label{single_b_SM_2q_diagrams}SM background single $b$ production.}
\end{center}
\end{figure}

Using FeynArts~\cite{FeynArts} and FormCalc~\cite{FormCalc}, which utilize FORM~\cite{FORM} and LoopTools~\cite{FormCalc}, we obtain the cross--sections presented in \citefigure{sigma_down_plot} and \citefigure{sigma_strange_plot}, with the numerical values for ${\sqrt{s}} = 10.58$~GeV given in \citetable{SM_qq_cross-section_values}.

\begin{table}[t]
\begin{center}
\begin{tabular}{|c|dr|}
\hline
$bd$ in SM & 7.3 & ${\times} 10^{-6}$~fb\\  
\hline                                      
$bs$ in SM & 1.8 & ${\times} 10^{-4}$~fb\\  
\hline                                      
\end{tabular}
\end{center}
\caption{\label{SM_qq_cross-section_values}SM background cross--sections for $e {\bar{e}} {\to} b {\bar{s}} / s {\bar{b}} / b {\bar{d}} / d {\bar{b}}$ at ${\sqrt{s}} = 10.58$~GeV.}
\end{table}

Considering the two--particle final states, the SM background is completely negligible compared to the squark--mediated signal for $b d$ production.  However, it is within an order of magnitude of the other three potential signals. Unfortunately, detecting such cross--sections of $10^{-4}$~fb is well beyond the reach of current colliders.

There are related processes, where four quarks are created in the hard process. They can then hadronize into two mesons, either a charged pair or a neutral pair.  The diagrams for the production of a charged pair are those in \citefigure{single_b_SM_4q_diagrams}.  Those for the production of a neutral pair are the same as for the charged pair, but with the down--type quarks combining to form a ${\bar{B}}^{0}$ and the up--types combining to form a light neutral meson.

\begin{figure}[t]
\begin{center}
\leavevmode
\includegraphics[width=12cm]{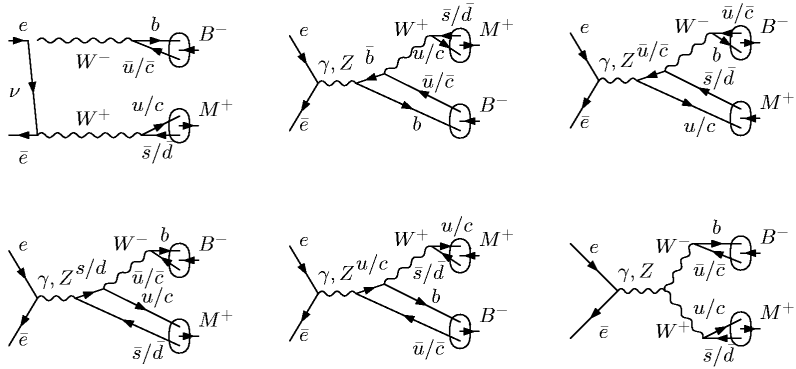}
\caption{\label{single_b_SM_4q_diagrams}Four--quark SM single $B$ meson production.}
\end{center}
\end{figure}

Generally, we expect the hard matrix element for the creation of four quarks to be of a similar size or less than the two--quark case.  Even ignoring the suppression of the wavefunction overlap of these four quarks with the two--meson final state, we can therefore safely neglect this Standard Model background as well.

\subsection{False Signal From ${\mathbf{B {\bar{B}}}}$ Pair Production}

Misidentification of $B$ mesons is an extremely important concern.  Our signal must not be confused with that of a $b {\bar{b}}$ pair production with one unidentified $b$.  Simply looking for events that contain only a single tagged bottom is (quantitatively) not feasible. Hence, we use kinematics to get rid of $b {\bar{b}}$ events.  The direct production of a $B$ meson and a light meson of mass $m_{M}$ leads to, in the beam center--of--mass frame, the $B$ meson taking a fraction $( s + m_{B}^{2} - m_{M}^{2} ) / (2 s)$ of the center--of--mass energy ${\sqrt{s}}$.  For the squark--mediated $bd$ signal with ${\sqrt{s}} = 10.58$~GeV, the $B$ meson will have energy between $6.56$~GeV (where the light meson is an ${{\eta}'}$) to $6.61$~GeV (where the light meson is a ${\pi}^{0}$).  This is to be compared to the case of $B {\bar{B}}$ production, where both have energy $5.29$~GeV.

The high--energy tail of the electron--positron beam can create $b {\bar{b}}$ pairs with enough energy that the resulting $B$ mesons could present a false signal by both having the energy that a singly--produced $B$ meson would have (around $6.6$~GeV for ${\sqrt{s}} = 10.58$~GeV), and one could decay into a high--energy light meson, with the radiated photon or particle missing the detector.  BaBar produces $1.1 {\times} 10^6$ $b {\bar{b}}$ pairs per fb${}^{-1}$, and has over $350$ fb${}^{-1}$ of integrated luminosity recorded~\cite{BaBar_data}.  This gives $385$~million $b {\bar{b}}$ pairs. The beam energy spread we expect to be of the order of $5$~MeV, estimated from the beam spread from $4.63$ to $4.83$~MeV on the ${\Upsilon}(4S)$ resonance~\cite{Upsilon_resonance_paper}.  For the false signal described, the $B {\bar{B}}$--pair is required to have $2.6$~GeV more than the mean beam energy.  This is over $400$ standard deviations away, if we assume that the beam enegy has a Gaussian distribution.  The expected number of events from this channel is then insignificant (less than $10^{-250}$).

Using the ${\Upsilon}(4S)$ resonance width of $20.7$~MeV~\cite{Upsilon_resonance_paper} as the spread, the cut is $125$ standard deviations away from the mean, which still leads to an expected number of events less than $10^{-250}$.  These brief estimates certainly allow us to neglect beam energy spread as a background source for our signal process.

This is also the source of any potential background to the case with an additional high--energy photon.  For the range of energies considered, the false signal background of a $B {\bar{B}}$ plus a high--energy photon requires between $1$ and $2$ GeV more than the mean beam energy.  This is $200$ to $400$ standard deviations over the mean, and hence the expected number of events is less than $10^{-250}$.  The false signal background of a $B {\bar{B}}$ pair of sufficient energy that the radiative decay of one of the mesons produces a photon that passes the cut is also less than $10^{-250}$ events.

\subsection{${\mathbf{R}}$--Parity Conserving MSSM Background}

\begin{figure}[t]
\begin{center}
\leavevmode
\includegraphics[width=10cm]{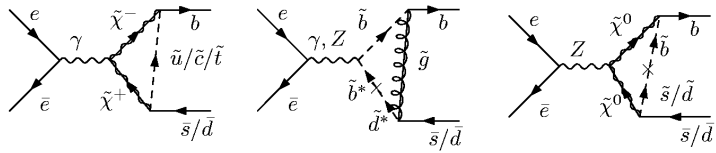}
\caption{\label{single_b_MSSM_diagrams}Example MSSM background diagrams: (a): flavor violation through $SU(2)_{L}$;  (b, c): flavor violation through a mass insertion on the squark line.  (b) is known as a \textit{pengluino} diagram.}
\end{center}
\end{figure}

Any signal of flavor violation in significant excess of the SM prediction is an exciting signal for new physics. However, the thrust of this paper is that such a signal could come from RPV couplings.  Backgrounds from the $R$--parity conserving part of the MSSM arise from two sources: flavor violation through $SU(2)_L$ and through non--minimal squark mixing, \ie~general soft SUSY breaking terms~\cite{non-minimal_QFV}. (Examples of both types are shown in \citefigure{single_b_MSSM_diagrams}.)

The diagrams for the former case are easily obtained by replacing the Standard Model particles in SM background loop diagrams with their supersymmetric partners.  The $W$ boson mass ($m_{W} {\gg} m_{B}$) accounts for most of the suppression of the SM background.  The sparticle masses are constrained to be (considerably) larger than $m_{W}$.  The structure of the amplitude is similar, which means that we can expect the SUSY loops without a new flavor structure to contribute below the level of the SM backgrounds. If we increase the largest sparticle mass in the loop to three times the $W$ boson mass, these SUSY backgrounds drop below $10\%$ to the already negligible Standard Model background rate.  There are potential enhancements in the large ${\tan} {\beta}$ region of the MSSM parameter space, but in the Higgs sector these destructively interfere with the SM amplitude~\cite{large_tan_beta_ref}, while any other enhancements are constrained by $b {\to} d {\gamma}$ to be at most close to the SM value.

The diagrams describing contributions from non--minimal flavor structure in squark sector are obtained by ``supersymmetrizing'' the virtual particles in the loops in the one--loop corrections to $e {\bar{e}} {\to} b {\bar{b}}$ (except for those diagrams without a virtual quark), and replacing the external ${\bar{b}}$ with a ${\bar{d}}$ and the internal ${\tilde{b}}$ with the mass eigenstate mixtures of ${\tilde{b}}$ and ${\tilde{d}}$.  These contributions are not easy to calculate, as the most significant pengluino diagram (shown in \citefigure{single_b_MSSM_diagrams}), is proportional to ${\alpha} {\alpha}_{s} {\delta} m_{{\tilde{q}}}^{2} / m_{{\tilde{g}}}^{2}$, where ${\delta} m_{{\tilde{q}}}^{2}$ is the difference in the squared masses of the squarks~\footnote{This assumes that the gluino is more massive than the squarks, otherwise replace the gluino mass with the mass of the more massive squark.}.  This, at least for $b$--$d$ mixing, is not well constrained~\cite{bs_mass_insertion_ref}.  However, we note that these diagrams would also contribute to $B {\to} {\rho} {\gamma}$, which \textit{is} tightly constrained.

Altogether, we expect the $R$--parity conserving part of the RPV MSSM to contribute to the background at a rate comparable to the Standard Model contribution at most.

\section{Outlook}

\begin{figure}[t]
\begin{center}
\leavevmode
\includegraphics[width=12cm]{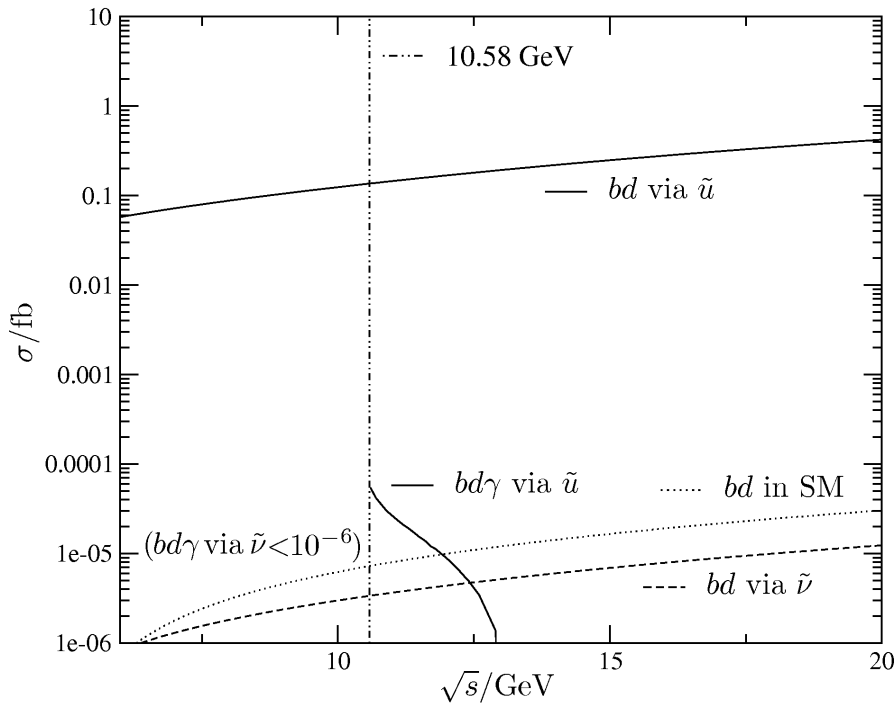}
\caption{\label{sigma_down_plot}Cross--sections for $e {\bar{e}} {\to} b {\bar{d}} / d {\bar{b}} / b {\bar{d}} {\gamma} / d {\bar{b}} {\gamma}$ through $R$--parity violation and the SM background for $e {\bar{e}} {\to} b {\bar{d}} / d {\bar{b}}$.}
\end{center}
\end{figure}

\begin{figure}[t]
\begin{center}
\leavevmode
\includegraphics[width=12cm]{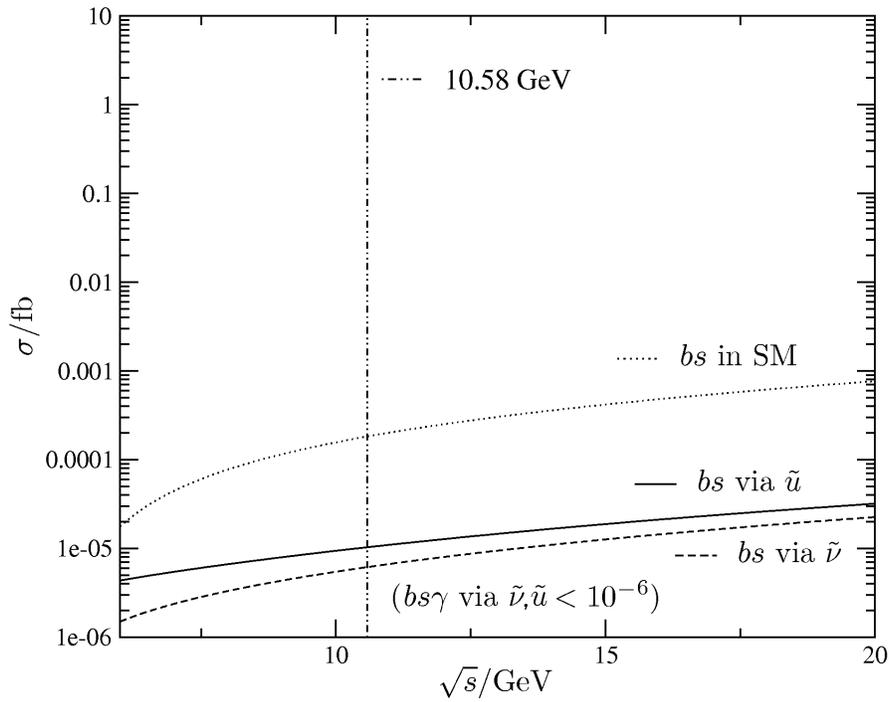}
\caption{\label{sigma_strange_plot}Cross--sections for $e {\bar{e}} {\to} b {\bar{s}} / s {\bar{b}} / b {\bar{s}} {\gamma} / s {\bar{b}} {\gamma}$ through $R$--parity violation and the SM background for $e {\bar{e}} {\to} b {\bar{s}} / s {\bar{b}}$.}
\end{center}
\end{figure}

As far as we are aware, there have been no searches for single $B$ production.  Currently BaBar has almost $400$~fb${}^{-1}$ of integrated luminosity~\cite{BaBar_data} and Belle has almost $650$~fb${}^{-1}$ of data~\cite{Belle_data} available for analyses.  Ignoring detector effects the maximum signal rate for single $b$ production allowed by current bounds comes from $t$--channel squark exchange and could be as large as $100$ events.

A null result, while disappointing, would still improve the bound on $|{{\lambda}'}_{1g1}|^{2} |{{\lambda}'}_{1g3}|^{2} m_{{\tilde{u}}}^{-4}$.  A $95\%$ confidence limit non--observation corresponds to $95\%$ confidence that less than three events occurred, leading to the deduction that the bound would be tightened by a factor of (expected events)$/2$, hence for a $0.13$~fb signal with $1$~ab${}^{-1}$ of luminosity, which is $130$ events with perfect detection efficiency, the bound on $|{{\lambda}'}_{1g1}|^{2} |{{\lambda}'}_{1g3}|^{2} m_{{\tilde{u}}}^{-4}$ would be tightened by a factor of $65$.

We have shown that the backgrounds to this process are negligible, which makes single $b$ production a promising search channel for $R$ parity violation.

\begin{acknowledgments}
I am grateful to Tilman Plehn, Michael Kr{\"{a}}mer, Thomas Binoth and Steve Playfer for useful discussions, and to Ulrich Nierste for reading the draft.  B.O'L.~is supported by the Carnegie Trust for the Universities of Scotland.
\end{acknowledgments}


\begin{thebibliography}{99}

\bibitem{MSSM_review} 
  for an introduction see \eg:
  S.~P.~Martin,
  arXiv:hep-ph/9709356;
  I.~J.~R.~Aitchison,
  arXiv:hep-ph/0505105.
\bibitem{RPV_review} 
  R.~Barbier {\it et al.},
  Phys.\ Rept.\  {\bf 420} (2005) 1
  [arXiv:hep-ph/0406039].
\bibitem{proton_decay_to_e_pi_ref} 
  M.~Shiozawa {\it et al.}  [Super-Kamiokande Collaboration],
  Phys.\ Rev.\ Lett.\  {\bf 81} (1998) 3319
  [arXiv:hep-ex/9806014].
\bibitem{neutron_decay_to_invisible_ref} 
  T.~Araki {\it et al.}  [KamLAND Collaboration],
  Phys.\ Rev.\ Lett.\  {\bf 96} (2006) 101802
  [arXiv:hep-ex/0512059].
\bibitem{PDG} 
  S. Eidelman {\it et al.},
  Phys.\ Lett.\ B {\bf 592}, 1 (2004)\\
  and 2005 partial update for the 2006 edition available on the PDG WWW pages (URL: http://pdg.lbl.gov/)
\bibitem{charged_current_universality} 
  see \eg:
  H.~K.~Dreiner, G.~Polesello and M.~Thormeier,
  Phys.\ Rev.\ D {\bf 65} (2002) 115006
  [arXiv:hep-ph/0112228].
\bibitem{my_RPV_bounds_paper} 
  H.~K.~Dreiner, M.~Kramer and B.~O'Leary,
  arXiv:hep-ph/0612278.
\bibitem{Saha_Kundu} 
  J.~P.~Saha and A.~Kundu,
  Phys.\ Rev.\ D {\bf 66} (2002) 054021
  [arXiv:hep-ph/0205046].
\bibitem{Xu_Wang_Yang} 
  Y.~G.~Xu, R.~M.~Wang and Y.~D.~Yang,
  Phys.\ Rev.\ D {\bf 74} (2006) 114019
  [arXiv:hep-ph/0610338].
\bibitem{APV_in_Cs} 
  J.~L.~Rosner,
  Phys.\ Rev.\ D {\bf 65} (2002) 073026
  [arXiv:hep-ph/0109239].
\bibitem{radiative_return} 
  G.~Rodrigo, H.~Czyz and J.~H.~Kuhn,
  eConf {\bf C0209101} (2002) WE06
  [Nucl.\ Phys.\ Proc.\ Suppl.\  {\bf 123} (2003) 167]
  [arXiv:hep-ph/0210287].
\bibitem{Lund_string} 
  B.~Andersson, G.~Gustafson, G.~Ingelman and T.~Sjostrand,
  Phys.\ Rept.\  {\bf 97} (1983) 31.
\bibitem{FeynArts} 
  T.~Hahn,
  Comput.\ Phys.\ Commun.\  {\bf 140} (2001) 418
  [arXiv:hep-ph/0012260].
\bibitem{FormCalc} 
  T.~Hahn and M.~Perez-Victoria,
  Comput.\ Phys.\ Commun.\  {\bf 118} (1999) 153
  [arXiv:hep-ph/9807565].
\bibitem{FORM} 
  J.~Vermaseren
  [arXiv:math-ph/0010025].
\bibitem{BaBar_data} 
  http://bbr-onlwww.slac.stanford.edu:8080/babarrc/perfdata.html
\bibitem{Upsilon_resonance_paper} 
  B.~Aubert {\it et al.}  [BABAR Collaboration],
  Phys.\ Rev.\ D {\bf 72} (2005) 032005
  [arXiv:hep-ex/0405025].
\bibitem{non-minimal_QFV} 
  L.~J.~Hall, V.~A.~Kostelecky and S.~Raby,
  Nucl.\ Phys.\ B {\bf 267} (1986) 415.
\bibitem{large_tan_beta_ref} 
  H.~E.~Logan and U.~Nierste,
  Nucl.\ Phys.\ B {\bf 586} (2000) 39
  [arXiv:hep-ph/0004139].
\bibitem{bs_mass_insertion_ref} 
  M.~B.~Causse and J.~Orloff,
  Eur.\ Phys.\ J.\ C {\bf 23} (2002) 749
  [arXiv:hep-ph/0012113].
\bibitem{Belle_data} 
  http://belle.kek.jp/bdocs/lum\_day.gif

\end{thebibliography}
\end{document}